\documentstyle[aps,prl,floats,epsf]{revtex}

\begin{document}
\draft

\twocolumn[\hsize\textwidth\columnwidth\hsize\csname@twocolumnfalse%
\endcsname

\title{Multi-Species Pair Annihilation Reactions}

\author{Olivier Deloubri\`ere {}$^1$, Henk J. Hilhorst {}$^2$, and
        Uwe C. T\"auber {}$^1$}

\address{$^1$ Department of Physics, Virginia Polytechnic Institute and 
              State University, Blacksburg, VA 24061-0435 \\
         $^2$ Laboratoire de Physique Th\'eorique, B\^atiment 210,
              Universit\'e de Paris-Sud, 91405 Orsay Cedex, France}
\date{\today} 
\vskip -0.4truecm

\maketitle

\begin{abstract}
We consider diffusion-limited reactions $A_i + A_j \to \emptyset$ 
($1 \leq i < j \leq q$) in $d$ space dimensions.
For $q > 2$ and $d \geq 2$ we argue that the asymptotic density decay for
such mutual annihilation processes with equal rates and initial densities
is the same as for single-species pair annihilation $A + A \to \emptyset$. 
In $d = 1$, however, particle segregation occurs for all $q < \infty$.
The total density decays according to a $q$ dependent power law,
$\rho(t) \sim t^{-\alpha(q)}$.
Within a simplified version of the model $\alpha(q) = (q-1) / 2q$ can be
determined exactly.
Our findings are supported through Monte Carlo simulations.
\end{abstract}
\vskip -0.2truecm
\pacs{PACS numbers: 05.40.-a, 82.20.Mj}
]

A large variety of systems in physics, chemistry, biology, and ecology 
can be modeled in terms of diffusion-limited reactions. 
This is because of their unifying feature of being composed of mobile 
agents (`particles') which interact upon encounter.
The traditional mean-field rate equations for such systems apply only to
homogeneous particle densities.
However, in many such systems the spatial dimension $d$ has one or more 
critical values below which density fluctuations invalidate the rate 
equations and new phenomena appear.
The fluctuations may stem from, {\em e.g.,} reaction-induced noise or 
initial state randomness, and typically dominate the system's large-scale, 
long-time behavior \cite{kuzovkov88}.
In order to extract the newly emerging nonequilibrium behavior, more 
sophisticated methods are needed:
extensive numerical simulations (for recent overviews, see 
Ref.~\cite{hinrichsen00}) along with powerful analytical methods such 
as scaling approaches; mappings to field theories followed by 
renormalization group; and exact solutions (mostly limited to $d=1$) 
\cite{chopard98}.

The theory of annihilation reactions has two landmarks, of which one, 
the single-species pair annihilation reaction $A + A \to \emptyset$, 
represents the simplest model case \cite{lee94}. 
The other one, the two-species annihilation $A + B \to \emptyset$, is 
considerably more subtle. 
It exhibits the remarkable phenomenon that for $d<4$ and for an
initially random particle distribution with equal densities 
$\rho_A(0) = \rho_B(0)$, the two species segregate into pure $A$ and 
pure $B$ domains, and the annihilations become localized within sharp 
reaction fronts between the domains \cite{lee95}.
It is a natural next step to ask for the long-time decay properties of a 
system of $q$ species that mutually annihilate according to
$A_i + A_j \to \emptyset$, with $1 \leq i < j \leq q$ \cite{benavraham86}.  
This `mutual annihilation model' (MAM) is the subject of this Letter
\cite{cardy02}.
A special case is the fully symmetric MAM, characterized by equal
reaction rates $\lambda_{ij}$, equal diffusion constants $D_i$, and
equal initial densities $\rho_i(0)$.
We will find that in $d \geq 2$ for all $2 < q < \infty$ the fully 
symmetric MAM behaves as the single-species pair annihilation process:
the total density $\rho(t) = \sum_i \rho_i(t) \sim t^{-1}$.
In stark contrast, in $d = 1$ it exhibits species segregation, and is
characterized by 
\begin{equation}
  \rho(t) \sim t^{-\alpha(q)} 
\label{density}
\end{equation}
with a $q$ dependent exponent given, within the approach presented 
below, by $\alpha(q)=(q-1) / 2q$.
We note that for $q \to \infty$, two particles of the same species meet 
with zero probability; the distinction between the different species 
then becomes irrelevant, and this model is equivalent to the 
$A + A \to \emptyset$ reaction \cite{krapivsky00}, with known 
$\alpha(\infty) = 1/2$.

In order to set the stage for our arguments, we briefly summarize the 
physics of the single- and two-species annihilation processes.
For $A + A \to \emptyset$ the mean-field rate equation 
$\dot{\rho}_A = - \lambda \rho_A^2$ with the solution 
$\rho_A(t) \sim 1 / \lambda t$ provides a valid description only for 
$d > 2$.
For $d < 2$ nearby reactant pairs quickly annihilate, leaving only
well-separated particles, which in turn slows the density decay down. 
These {\em anti}-correlations mimic a repulsion between the particles; 
in a field theory representation of the associated master equation
\cite{doi76} they lead to a downward renormalization of $\lambda$.
As the diffusion propagator remains unchanged, the perturbation series
is readily summed to all orders; one finds $\rho_A(t) \sim t^{-d/2}$ 
for $d < 2$ and $\rho_A(t) \sim t^{-1} \ln t$ at the critical 
dimension $d_c = 2$ \cite{lee94}.

For the two-species pair annihilation $A+B \to \emptyset$ the rate 
equations read $\dot{\rho}_{A/B} = - \lambda_{AB} \rho_A \rho_B$.
With equal initial densities $\rho_A(0) = \rho_B(0)$ they are solved
again by $\rho_{A/B}(t) \sim 1/t$;
with $\rho_A(0) > \rho_B(0)$, say, one obtains 
$\rho_B(t) \sim \exp(- \lambda_{AB} [\rho_A(0) - \rho_B(0)] t)$ for the
minority species, while the majority approaches a saturation density 
$\rho_A(\infty)$.
In order to establish the effects of spatial fluctuations, it is crucial
to notice that the density difference $\rho_A - \rho_B$ remains conserved 
under the reactions; for $D_A = D_B$ it simply obeys the diffusion 
equation \cite{toussaint83}.
Initial density difference fluctuations therefore amplify in time 
relative to the total density.
As a consequence, when $\rho_A(0)=\rho_B(0)$, then for $d < d_c=4$  
the system segregates into domains and the 
density decay is slowed down to $\rho_{A/B}(t) \sim t^{-d/4}$ 
\cite{bramson91}. 
The renormalization group provides a firm basis \cite{lee95} for these 
arguments, at least for $2 \leq d < 4$:
An effective field theory can be derived that corresponds to the rate 
equations plus diffusion terms, which establishes the segregation.

For unequal diffusion constants, $D_A \not= D_B$, this picture is not
qualitatively altered; however, special initial conditions may change it. 
Consider, {\em e.g.,} particles that initially alternate in 
one-dimensional space, $ABAB \ldots$, and that upon encounter react with 
probability one. 
The distinction between $A$ and $B$ is then meaningless and the system 
is in the $A + A \to \emptyset$ universality class \cite{krapivsky00}.
For unbalanced initial conditions, $\rho_A(0) > \rho_B(0)$, stretched
\cite{kang84} rather than simple exponential relaxation ensues for 
$d < 2$: $\ln \rho_B(t) \sim - t^{d/2}$,  whereas 
$\ln \rho_B(t) \sim - t / \ln t$ at $d_c = 2$ \cite{bramson91}.
If the two species are already segregated initially, $d_c = 2$ also is 
the sole critical dimension \cite{cornell93}.

This summary helps us classify the possible scenarios for the 
$q$-species MAM with arbitrary parameters $\lambda_{ij},\, D_i$, and
$\rho_i(0)$. 
Generically we expect that after some crossover time only the least
reactive and initially most numerous species will have survived, 
resulting in an effective two-species problem.
After this reduction of $q$ to the effective value $q_{\rm eff} = 2$ the 
final asymptotic decay laws will be those of the two-species system with 
unequal initial densities discussed above.
However, on special submanifolds of parameter space, and in particular 
in the presence of symmetries, reduction to $q_{\rm eff} = 2$ may not be 
possible and novel behavior may appear. 
That not {\em all} symmetries lead to new behavior may be illustrated by 
the cyclic reaction scheme $A + B \to \emptyset$, $B + C \to \emptyset$, 
$C + D \to \emptyset$, and $D + A \to \emptyset$, all with equal rates 
and initial densities.
Here, we may readily identify the species $A \equiv C$ and $B \equiv D$, 
respectively, which takes us back to the $A + B \to \emptyset$ reaction
with $\rho_A(0) = \rho_B(0)$.
In this Letter we address the most prominent case that requires special 
consideration, and will in fact lead to novel effects, namely the fully 
symmetric MAM, in which all $q$ species are equivalent (whence 
$q_{\rm eff} = q$). 

First, we notice that the renormalization of the annihilation vertices 
in this $q$-species model is independent of $q$ and identical to that 
of the $A + A \to \emptyset$ reaction \cite{lee95}, with $d_c = 2$ 
\cite{lee94}.
Second, for $q > 2$ there exists no microscopic, local conservation law.
As a consequence, following the arguments in Ref.~\cite{lee95}, any
memory of the initial state will eventually become lost.
This eliminates the segregation mechanism at work in the $q = 2$ case.
Next we invoke the mean-field rate equations and conclude that
$\rho_i(t) = \rho(t) / q \sim 1/t$ for $d > 2$ \cite{remark1}.

For $d \leq 2$, however, one needs to extract the correct asymptotic 
scaling from the Callan--Symanzik renormalization group equation.
This requires an explicit computation of the density (a function of its 
initial value) as a power series in the renormalized annihilation rate 
$\lambda_R$.
At the critical dimension $d_c = 2$ the renormalized rate flows to zero 
logarithmically, $\lambda_R(t) \sim 1 / \ln t$, leaving merely the tree 
diagram contributions that correspond to the solution of the rate 
equation.
Thus we predict $\rho_i(t) = \rho(t) / q \sim t^{-1} \ln t$ in $d = 2$.
The difficulty for $d < 2$ is to demonstrate that for large values of the 
relevant operator $\rho(0)$ the power series remains properly controlled 
\cite{lee94}.
For $d < 2$ this has proven elusive even for the two-species system
\cite{lee95}. 
Moreover, in one space dimension blockage effects of (hard-core) 
particles have been found to often play a crucial role in multi-species
reaction-diffusion processes \cite{odor01}.
We shall see that the $q$-species MAM, too, develops entirely novel 
features when restricted to a linear chain:
The $q$ species segregate into well-defined domains, which remain stable 
because of the mutual annihilation processes that prevent species mixing
and the special one-dimensional topological constraints that do not allow 
a given species to interact with all others.
As a consequence, we find that 
for all $2 \leq q < \infty$ that the total 
density decays in $d=1$ according to the power law (\ref{density}).
\begin{figure}[t]
\epsfxsize=0.95\columnwidth \epsfbox{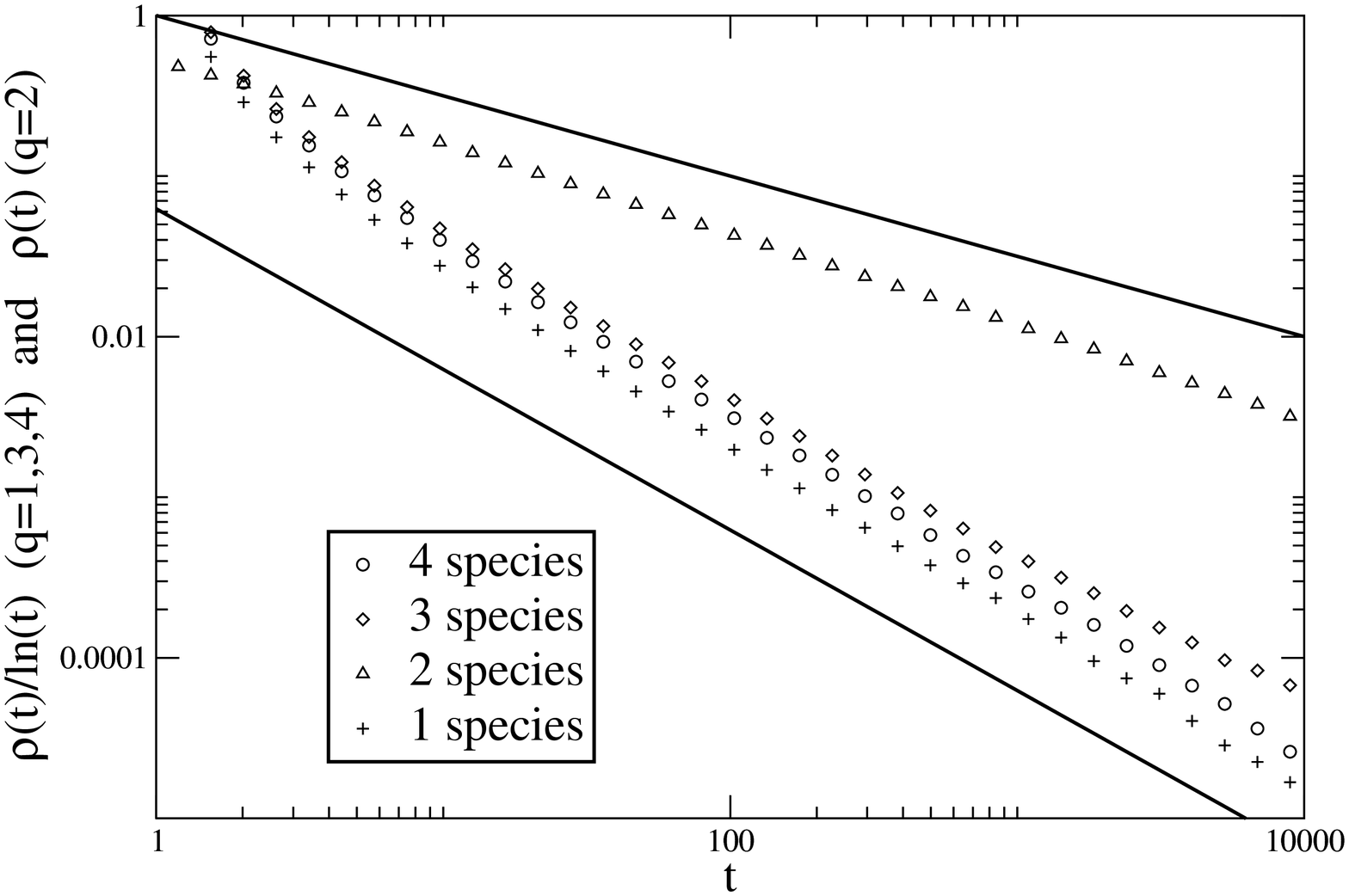}
\caption{Monte Carlo results for the total density decay 
   vs. time in the pair annihilation reactions $A + A \to \emptyset$ and 
   $A_i + A_j \to \emptyset$ with $1 \leq i < j \leq q$ ($q = 2,3,4$) 
   for equal initial densities in two dimensions.
   The plots for $q \not= 2$ depict $\rho(t) / \ln t$; the solid lines 
   indicate the functions $t^{-1}$ and $t^{-1/2}$.}
\label{2ddecay}
\vskip -0.3truecm
\end{figure}

We now present our numerical evidence in $d = 1$ and $d = 2$, and then 
proceed with the analysis of the one-dimensional model.
In order to check the predicted universal decay law in two dimensions, 
we performed Monte Carlo simulations on a $512 \times 512$ square 
lattice with hard-core particles.
Starting from a full lattice with random distribution of $q$ equally
abundant species ($q = 2,3,4$), we let the particles perform unbiased 
random walks and annihilate with probability one upon encounter with a 
different species.
One Monte Carlo time step was considered complete after $N(t)$ trials, 
with $N(t)$ the number of remaining particles at that instant.
The results, shown in Fig.~\ref{2ddecay}, clearly support 
$\rho(t) \sim t^{-1} \ln t$ for the $q$-species MAM with $q = 3,4$;
this is similar to the decay law of the $A + A \to \emptyset$ reaction, 
and in contradistinction to $\rho(t) \sim t^{-1/2}$ for $q = 2$.
We have also checked the purely mean-field behavior for $q = 3,4$ in a 
$50^3$ cubic lattice \cite{deloubriere03}.

\begin{figure}[t]
\epsfxsize=0.95\columnwidth \epsfbox{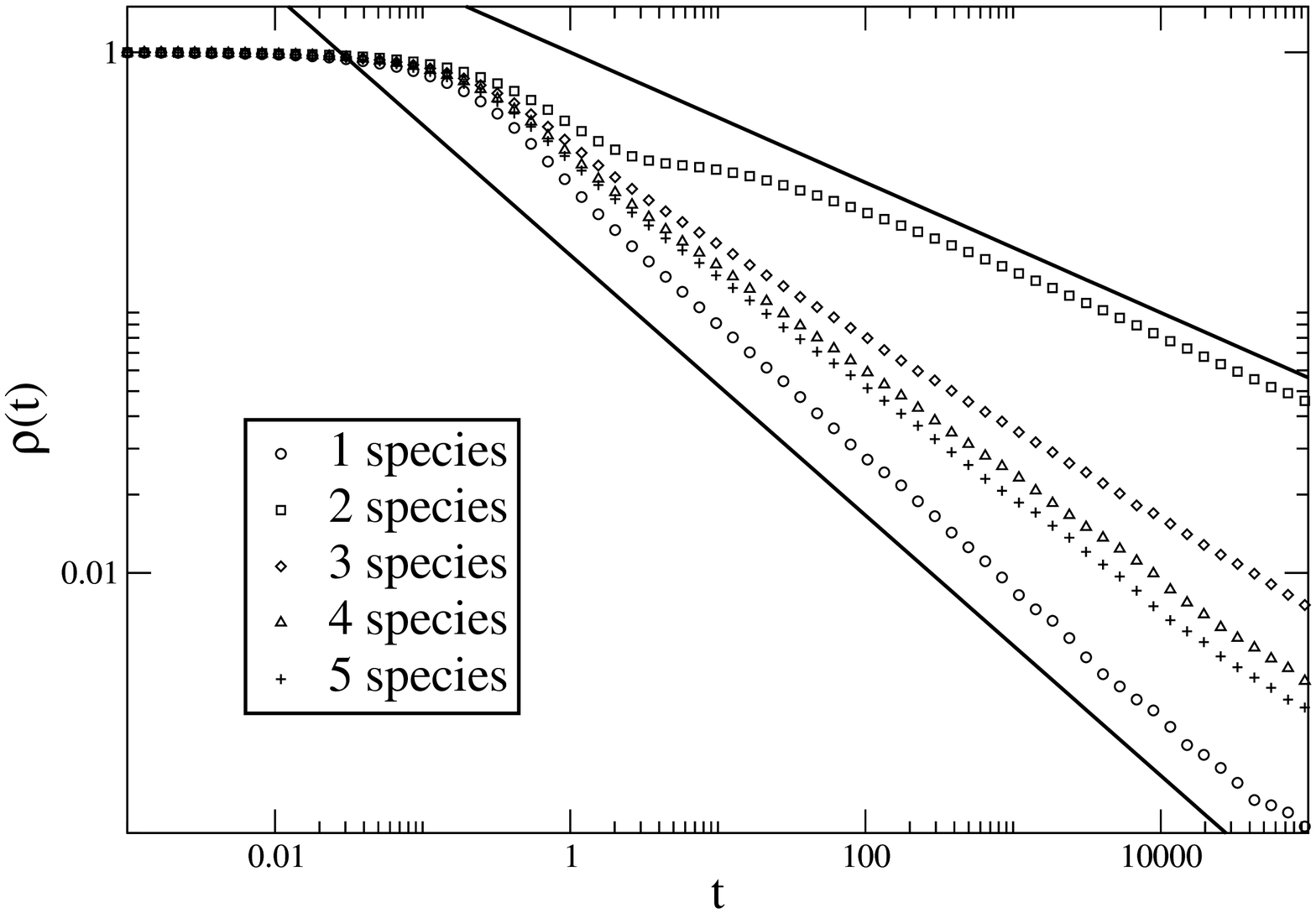}
\caption{Monte Carlo results for the total density decay 
   $\rho(t)$ vs. $t$ in the pair annihilation reactions 
   $A + A \to \emptyset$ and $A_i + A_j \to \emptyset$ with 
   $1 \leq i < j \leq q$ ($q = 2,3,4,5$) and equal initial densities in 
   one dimension.
   The solid lines indicate the functions $t^{-1/2}$ and $t^{-1/4}$.}
\label{1ddecay}
\vskip -0.3truecm
\end{figure}

In $d = 1$, however, simulations of the MAM with equal initial densities 
yield decay laws that differ importantly from both the 
$A + A \to \emptyset$ and $A + B \to \emptyset$ cases \cite{krapivsky00}.
Figure~\ref{1ddecay} shows our Monte Carlo results for $q = 2,3,4,5$ on 
a chain of $10^5$ sites with periodic boundary conditions.
Evidently at long times $\rho(t) \sim t^{-\alpha(q)}$, where $\alpha(q)$ 
increases with $q$ from $\alpha(2) = 1/4$ (the $A + B \to \emptyset$ 
value) to $\alpha(\infty) = 1/2$ (the $A + A \to \emptyset$ value).

In order to study the one-dimensional MAM we simplify it such as to 
retain only its barest essentials. 
This simplified version, to be referred to as SMAM, arises from the 
following considerations. 
The one-dimensional system may at any time be decomposed into a sequence 
of domains, each containing only a single particle species.
Owing to the diffusive nature of the process the typical domain size
increases as $L(t) \sim (Dt)^{1/2}$. 
Let us assume the asymptotic decay law
$\rho(t) \sim \rho_0 (\rho_0^2 Dt)^{-\alpha}$, where $\alpha$ remains to 
be determined.
The average particle number in a domain then scales as 
$\overline{n}(t) = L(t) \rho(t) \sim (\rho_0^2 Dt)^{- \alpha + 1/2}$,
and phase segregation occurs only if $\alpha < 1/2$.
Adjacent domains are separated by reaction zones, of which there are 
$1/L(t)$ per unit of length.
Therefore, as argued in Ref.~\cite{leyvraz92} for the two-species case,
the total particle density decreases as 
$\dot{\rho}(t) = - \kappa(t) / L(t)$, with $\kappa(t)$ the typical 
number of annihilations per unit of time in a zone. 
The SMAM is now defined by the assertion that fluctuations in the 
annihilation rate $\kappa(t)$, whether in the course of time or between 
different reaction zones, are irrelevant and may be ignored; {\em i.e.,} 
the particle content of each domain, owing to the annihilations taking 
place at both of its ends, decreases at the uniform rate $2 \kappa(t)$.
To complete the picture, we need to specify what happens when a domain 
loses all its particles:
Then, with probability $1/(q-1)$, the left and right neighboring domains 
contain identical species, and consequently fuse into a single new 
domain; or, with the complementary probability $(q-2)/(q-1)$, they 
contain different particle species and a new reaction zone appears.

The SMAM may be cast into an explicit algorithm.
We consider a one-dimensional lattice whose sites 
$1 \leq i \leq N^{(0)}$ represent the {\em domains} of the original MAM.
We randomly select initial values $n_i^{(0)}$ for the particle numbers 
in each domain.
This random initial state evolves in time via {\em deterministic} 
iterations.
The $(k+1)$th iteration changes the total number of sites from 
$N^{(k)}$ to $N^{(k+1)}$ and converts the integer set 
$\{ n_i^{(k)} \}_{i=1}^{N^{(k)}}$ to $\{ n_j^{(k+1)} \}_{j=1}^{N^{(k+1)}}$ 
by successive application of the following four iteration steps:
(i) All $n_i^{(k)}$ are reduced by $1$.
(ii) All sites that as a result have become empty,
are eliminated and the other sites are reconnected without reordering.
(iii) Any two sites that as a result have become neighbors, are, with 
probability $1/(q-1)$, fused into a single site whose number variable is 
the sum of the number variables of the fusing sites.
(iv) The remaining sites are relabeled with an index 
$1 \leq j \leq N^{(k+1)}$.
The $k$th iteration yields the total number of domains $N^{(k)}$ and the
average particle number $\overline{n}^{(k)}$ per domain.
The particle density and the physical time follow from 
$\rho^{(k)} / \rho^{(0)} = N^{(k)} \overline{n}^{(k)} / N^{(0)} 
\overline{n}^{(0)}$, and $t^{(k)} / t^{(0)} = [ N^{(0)} / N^{(k)} ]^2$,
as $N(t) \sim L(t)^{-1} \sim t^{-1/2}$.

A key feature of the SMAM is that at every iteration step $k$ the numbers
$n_i^{(k)}$ are uncorrelated, for they descend from disjoint sets of 
`ancestor' variables.
Therefore this model obeys an exact closed set of equations, which we 
will now derive and analyze.
Let us for the moment suppress the iteration superscript $(k)$, and 
denote, preceding the $k$th iteration, the total number of domains by $N$, 
the total number of domains containing $n$ particles by $M_n$, and their
relative abundance by $f_n = M_n / N$.
Primed symbols will indicate the corresponding quantities after the 
$k$th iteration. 
In one iteration the total number $N$ of domains diminishes by $M_1$ due
to step (i). 
Step (iii) results in the additional disappearance of domains; calculating 
their exact number requires taking into account all instances where two or 
more vanishing domains form a sequence of nearest or next-nearest 
neighbors.
After some combinatorics one finds \cite{deloubriere03} that $N'$ and $N$ 
are related by $N' = (1 - f_1) \, [1 - f_1 / (q-1)] N$.
By means of more elaborate combinatorial analysis one may express the 
final number $M_n'$ of $n$ particle domains in terms of the initial $M_m$.
The intensive variables $f_n$ ($n=1,2,\ldots$) then obey the recursion 
relation
\begin{equation}
  f_n' = [1 + (q-2) \tilde{f}_1] \, (f_{n+1} + \tilde{f}_1 {\cal R}_n) \ ,
\label{fullrecursion}
\end{equation}
with the abbreviations $\tilde{f}_1 = f_1 /[(q-1)(1-f_1)]$ and 
${\cal R}_n = \sum_{s=2}^\infty \tilde{f}_1^{s-2} 
\sum_{m_1,\ldots,m_s=1}^\infty f_{m_1+1} \ldots f_{m_s+1} \, 
\delta_{n,m_1+\ldots+m_s}$.
The term of index $s$ represents the creation of a domain of $n$ 
particles by simultaneous fusion of $s$ domains. 
The fusions with $s \geq 3$ are clearly very model specific and one would 
expect the essential physics to be embodied already in the lowest-order 
nonlinearity. 
Indeed, by truncating Eq.~(\ref{fullrecursion}) after the $s=2$ term one 
obtains an elegant Boltzmann-like equation; the mathematical analysis 
below is easier, however, if all terms are retained.

To find a solution to Eq.~(\ref{fullrecursion}) we substitute an 
exponential distribution $f_n = \epsilon (1 - \epsilon)^{n-1}$.
The recursion then yields a similar distribution, but with a new
parameter $\epsilon'$ related to $\epsilon$ by
$\epsilon' = \epsilon [1 - \epsilon / (q-1)]$.
For this solution $f_1 = \epsilon$, which may be substituted in the
relation linking $N'$ to $N$. 
Since $\overline{n} = 1/\epsilon$, the total particle density obeys
$\rho' / \rho = N' \epsilon / N \epsilon' = 1 - \epsilon$.
After restoring the iteration indices we obtain the pair of recursion
relations
\begin{eqnarray}
\epsilon^{(k+1)}&=&\epsilon^{(k)}[1-\epsilon^{(k)}/(q-1)] \ ,
\label{epsrecursion}\\[2mm]
\rho^{(k+1)}&=&\rho^{(k)}[1-\epsilon^{(k)}] \ ,
\label{rhorecursion}
\end{eqnarray}
to be solved with initial condition $0 < \epsilon^{(0)} < 1$ 
[{\em e.g.,} for a random initial distribution $\epsilon^{(0)} = (q-1)/q$] 
and $\rho^{(0)}$.
The solution of Eqs.~(\ref{epsrecursion}) and (\ref{rhorecursion})
determines $\rho^{(k)}$ and 
$t^{(k)}=t^{(0)}[\rho^{(0)}\epsilon^{(0)}/\rho^{(k)}\epsilon^{(k)}]^2$;
the desired function $\rho(t)$ is then obtained by eliminating the index 
$k$.

We have been able to carry this through explicitly only in an asymptotic 
expansion for large $k$. 
Whereas its leading order is readily evaluated, more attention is 
required to deal with the subleading correction.
By analyzing the recursion relation (\ref{epsrecursion}) one finds that
\begin{equation}
  \epsilon^{(k)} = \frac{q-1}{k} \left[ 1 - \frac{\log ck}{k} 
  + {\cal O}\left( \frac{\log^2k}{k^2} \right) \right] \ ,
\label{epsasptk}
\end{equation}
where $c$ is a function of $\epsilon^{(0)}/(q-1)$.
Analyzing Eq.\,(\ref{rhorecursion}) with (\ref{epsasptk}) inserted then 
yields
\begin{equation}
  \rho^{(k)} \simeq \frac{A \, \rho^{(0)}}{k^{q-1}} 
  \left[ 1 - (q-1) \frac{\log c k - (q-2)/2}{k} \right]
\label{rhoasptk}
\end{equation}
with $A = \lim_{k \to \infty} k^{q-1} \prod_{\ell=0}^{k-1} 
[1 - \epsilon^{(\ell)}]$.
Expressing $t^{(k)}$ as a function of $k$ and inverting leads to
$k(t) \simeq (C t)^{1/2q} - [\log(C t) + 2q \log c - (q-1)(q-2)] / 2 q$
with $C = A^2 (q-1)^2 /$ ${\epsilon^{(0)}}^2 t^{(0)}$.
Finally, upon substitution in Eq.~(\ref{rhoasptk}), 
\begin{equation}
  \rho(t) \simeq A \rho^{(0)} \left[ (Ct)^{-\alpha(q)} + 
  \frac{(q-1)(q-2)}{2q} \, (Ct)^{-1/2} \right]
\label{rhoasptt}
\end{equation}
with $\alpha(q) = 1/2 - 1/2q < 1/2$, confirming species segregation
self-consistently and establishing Eq.~(\ref{density}) for the leading 
density decay of the SMAM.
For $q = 2$ we recover $\alpha(2) = 1/4$, and the limit $q \to \infty$
gives correctly $\alpha(\infty) = 1/2$.
Notice that the term with $\log(Ct)$ and the dependence on 
$c$ have canceled out in Eq.~(\ref{rhoasptt}).
The next-to-leading behavior is identical with the power law decay for 
the $A + A \to \emptyset$ reaction in $d = 1$.
Its relative amplitude increases with $q$; thus it becomes numerically
difficult to distinguish it from the leading term.
We cannot establish that the correction term in Eq.~(\ref{rhoasptt}) 
has the same relevance for the original MAM as we believe the 
leading-order term does; in the accessible time window of the MAM 
simulations our data are best described by an effective exponent 
$\alpha_{\rm eff}$, which reflects a sizeable next-to-leading correction 
\cite{deloubriere03}. 
Current large-scale simulations by Ben-Avraham and Zhong indeed confirm 
unambiguously both our leading density decay law (\ref{density}) as well
as the power $t^{-1/2}$ for the first correction term \cite{benavraham02}.

In this one-dimensional system the reaction zone width $\ell(t)$ is just
equal to the
typical interparticle distance between representatives of different 
species.
The reaction rate $\kappa(t)$ is just the inverse of the time needed to 
diffuse over this length \cite{leyvraz92}, hence 
$\kappa(t) \sim D / \ell(t)^2$.
Combining this with $\dot{\rho}=-\kappa/L$ and the known time dependences
of $L(t)$ and $\rho(t)$, we find 
$\ell(t)^2 \sim \rho_0^{-2} (\rho_0^2 D t)^{\alpha(q)+1/2}$, whence
\begin{equation}
  \ell(t) \sim t^{\lambda(q)} 
\label{exponentrel}
\end{equation}
with $\lambda(q) =(2q-1) / 4 q$.
For $q=2$ this reproduces the known result $\lambda(2) = 3/8$ 
\cite{leyvraz92}. 
The value $\lambda(\infty) = 1/2$ indicates that for infinitely many 
species the reaction zones grow as fast as the typical domain size, and 
hence there can be no segregation.
How to aptly take into account the special one-dimensional topological 
restrictions in a field-theoretic description remains an open issue.

This research has in part been supported by the NSF (grant DMR-0075725)
and the Jeffress Memorial Trust (J-594).
We thank Paul Krapivsky for bringing this problem to our attention, and 
gladly acknowledge helpful discussions with Dani ben-Avraham, Sid Redner, 
Jaime Santos, Beate Schmittmann, Fr\'ed\'eric van Wijland,
Ben Vollmayr-Lee, and Royce Zia.

\end{document}